\documentclass[reprint,amsfonts, amssymb, amsmath,  showkeys,pra, superscriptaddress, twocolumn,longbibliography,nofootinbib]{revtex4-2}

\usepackage{xcolor}

\usepackage{float}
\makeatletter
\let\newfloat\newfloat@ltx
\makeatother
\usepackage[english]{babel}
\usepackage[utf8]{inputenc}
\usepackage{graphics}
\usepackage{selinput}
\usepackage[normalem]{ulem}
\usepackage[shortlabels]{enumitem}

\usepackage{braket}
\usepackage{amsthm}
\usepackage{mathtools}
\usepackage{physics}
\usepackage{graphicx}
\usepackage[left=16mm,right=16mm,top=35mm,columnsep=15pt]{geometry} 
\usepackage{adjustbox}
\usepackage{placeins}
\usepackage[T1]{fontenc}
\usepackage{lipsum}
\usepackage{csquotes}
\usepackage{bm}

\usepackage[linesnumbered,ruled,vlined]{algorithm2e}
\SetKwInput{kwInit}{Init}

\def\HC{\mathcal{H}}

\def\LC{\mathcal{L}}

\def\ad{^{\dagger}}

\newcommand{\fsnull}[1]{}
\newcommand{\old}[1]{}

\usepackage[makeroom]{cancel}
\usepackage[toc,page]{appendix}
\usepackage[colorlinks=true,citecolor=blue,linkcolor=magenta]{hyperref}

\usepackage{tikz}
\tikzset{every picture/.style=remember picture}

\usepackage[utf8]{inputenc}
\usepackage{graphicx}
\usepackage{xcolor}
\usepackage{amsmath}
\usepackage{amsthm}
\usepackage{bm}
\usepackage{bbm}
\usepackage{comment}
\usepackage{appendix}
\usepackage{mathdots}
\usepackage{lipsum}
\usepackage{verbatim}
\usepackage{natbib}
\usepackage{nccmath}
\usepackage{amsfonts} 

\newcommand{\dya}[1]{\ket{#1}\!\bra{#1}}

\newcommand{\AC}{\mathcal{A}}
\newcommand{\BC}{\mathcal{B}}

\newcommand{\EC}{\mathcal{E}}
\newcommand{\FC}{\mathcal{F}}
\newcommand{\GC}{\mathcal{G}}

\newcommand{\PC}{\mathcal{P}}

\newcommand{\SC}{\mathcal{S}}

\renewcommand{\geq}{\geqslant}
\renewcommand{\leq}{\leqslant}

\newcommand*{\id}{\openone}

\newcommand{\bs}{\textsf{BS}}

\def\calF{\mathcal{F}}

\def\calP{\mathcal{P}}

\def\be{\begin{equation}}
\def\ee{\end{equation}}
\def\bs{\begin{split}}
\def\e{\end{split}}
\def\ba{\begin{eqnarray}}
\def\bea{\begin{eqnarray}}

\def\tea{\end{eqnarray}}
\def\ea{\end{eqnarray}}
\def\eea{\end{eqnarray}}

\newcommand\mbb[1]{\mathbb{#1}}

\newtheorem{theorem}{Theorem}

\usepackage{amssymb}
\usepackage{dsfont}

\def\be{\begin{equation}}
\def\te{\end{equation}}
\def\ee{\end{equation}}
\def\ba{\begin{eqnarray}}
\def\bea{\begin{eqnarray}}

\def\tea{\end{eqnarray}}
\def\ea{\end{eqnarray}}
\def\eea{\end{eqnarray}}

\def\GL{\mathbb{GL}}

\begin{document}

\title{A unified approach to quantum resource theories and a new class of free operations}

\author{N. L. Diaz}
\affiliation{Information Sciences, Los Alamos National Laboratory, Los Alamos, NM 87545, USA}
\affiliation{Center for Non-Linear Studies, Los Alamos National Laboratory, 87545 NM, USA}

\author{Antonio Anna Mele}
\affiliation{Dahlem Center for Complex Quantum Systems, Freie Universität Berlin, 14195 Berlin, Germany}
\affiliation{Theoretical Division, Los Alamos National Laboratory, Los Alamos, NM 87545, USA}

\author{Pablo Bermejo}
\affiliation{Information Sciences, Los Alamos National Laboratory, Los Alamos, NM 87545, USA}
\affiliation{Donostia International Physics Center, Paseo Manuel de Lardizabal 4, E-20018 San Sebasti\'an, Spain}
\affiliation{Department of Applied Physics, University of the Basque
Country (UPV/EHU), 20018 San Sebastián, Spain}

\author{\\Paolo Braccia}
\affiliation{Theoretical Division, Los Alamos National Laboratory, Los Alamos, NM 87545, USA}

\author{Andrew E. Deneris}
\affiliation{Information Sciences, Los Alamos National Laboratory, Los Alamos, NM 87545, USA}
\affiliation{Quantum Science Center, Oak Ridge, TN 37931, USA}

\author{Mart\'{i}n Larocca}
\affiliation{Theoretical Division, Los Alamos National Laboratory, Los Alamos, NM 87545, USA}

\author{M. Cerezo}
\thanks{cerezo@lanl.gov}
\affiliation{Information Sciences, Los Alamos National Laboratory, Los Alamos, NM 87545, USA}
\affiliation{Quantum Science Center, Oak Ridge, TN 37931, USA}

\begin{abstract}
In quantum resource theories (QRTs) certain quantum states and operations are deemed more valuable than others. While the determination of the ``free'' elements is usually guided by the constraints of some experimental setup, this can make it difficult to study similarities and differences between QRTs. In this work, we argue that QRTs follow from the choice of a preferred algebraic structure $\mathcal{E}$  to be preserved, thus setting the free operations as the automorphisms of $\mathcal{E}$. We illustrate our finding by determining $\mathcal{E}$ for the QRTs of entanglement, Clifford stabilizerness, purity, imaginarity, fermionic Gaussianity,  reference frames, thermodynamics and coherence; showing instances where $\mathcal{E}$ is a Lie algebra, group, ring, or even a simple set. This unified understanding allows us to generalize the concept of stochastic local operations and classical communication (SLOCC) to identify novel resource non-increasing operations for Lie-algebra based QRTs, thus finding a new solution to an open problem in the literature. We showcase  the sanity of our new set of operations by rigorously proving that they map free states to free states, as well as determine more general situations where these transformations strictly do not increase the resource of a state.
\end{abstract}

\maketitle

\textbf{Introduction.}  Resource Theories (RTs) acknowledge the fact that objects within a system can acquire value depending on how easy, or hard, it is to obtain them~\cite{costa2020information,coecke2016mathematical}. As such, they provide a mathematical  framework  to systematically study, quantify and manipulate these resources.  At their core, RTs can be defined  in terms of two fundamental notions. First, one determines a restricted set of \textit{free operations} which are assumed to not generate any resource, and which represent the limitations or constraints within the system. Then, one defines the set of \textit{free states} of the system as those  that can be prepared via free operations at no additional cost. From here, \textit{resourceful operations and states} are defined  as those which are not free. Indeed, the term ``resourceful'' in an RT usually  carries operational meaning as resourceful states can be used to accomplish operations and tasks which would be impossible with only free states and operations.

The application of RTs to quantum mechanical systems has led to the development  of Quantum Resource Theories (QRTs)~\cite{chitambar2019quantum}, which can become a central tool in quantum information sciences. For instance, one can define QRTs for entanglement~\cite{horodecki2009quantum,plenio2005introduction,beckey2021computable} and extensions thereof~\cite{contreras2019resource,kaur2021resource}, scrambling~\cite{garcia2023resource}, Clifford stabilizerness~\cite{veitch2014resource,howard2017application,leone2024stabilizer}, 
purity~\cite{horodecki2003reversible,streltsov2018maximal,gour2015resource,luo2019robustness}, imaginarity~\cite{hickey2018quantifying,wu2021resource,wu2021operational},  fermionic Gaussianity~\cite{gigena2015entanglement,weedbrook2012gaussian,hebenstreit2019all,diaz2023showcasing,jozsa2008matchgates,mele2024efficient,dias2023classical,goh2023lie,brod2011extending}, spin coherence~\cite{robert2021coherent,perelomov1977generalized,zhang1990coherent}, reference frames~\cite{vaccaro2008tradeoff,gour2008resource,marvian2013theory,marvian2016quantum,martinelli2019quantifying}, thermodynamics~\cite{gour2015resource,goold2016role,lostaglio2019introductory,ng2019resource}, and coherence~\cite{streltsov2017colloquium,saxena2020dynamical,wu2020quantum,winter2016operational,smith2017quantifying}. 

In general, the free states and operations in QRTs are determined from experimental constraints. For instance, in the theory of entanglement, two spatially separated parties can prepare and act on local states from a joint quantum system, as well as use classical randomness (e.g., throw dice) and send each other messages over classical channels (e.g., make phone calls). These choices, while operationally motivated, can sometimes seem arbitrary and make it hard to extrapolate known concepts from one QRT onto another, as well as lead to gaps between operational and axiomatic definitions of free operations~\cite{heimendahl2022axiomatic,koashi2008quantum,duan2009distinguishability}. This raises  
questions like: \textit{How do we fully characterize resource non-increasing operations in general QRTs, such as that of fermionic Gaussianity? What is the equivalent of phone calls in a fermionic system?} This issue has been reported in the literature (see, e.g., Ref.~\cite{barnum2003generalizations}), and while several approaches have been proposed ~\cite{veitch2014resource}, these are strongly inspired by the QRT of entanglement and can be heavily biased towards exploiting an underlying tensor product structure of the Hilbert space. Thus, one may wonder if the key to answering these questions lies in understanding and unifying QRTs from a more fundamental perspective. One such intriguing approach can be traced back to Refs.~\cite{barnum2003generalizations,barnum2004subsystem,viola2007generalized,weinstein2006generalized}, where it was shown that some QRTs can be defined from, and relative to, a preferred set of observables. While these ideas can be used as a unifying principle, the proposed framework has several shortcomings such as failing in scenarios where the set of preferred operators is not a Lie algebra. 

\begin{figure}[t!]
    \centering
    \includegraphics[width=1\linewidth]{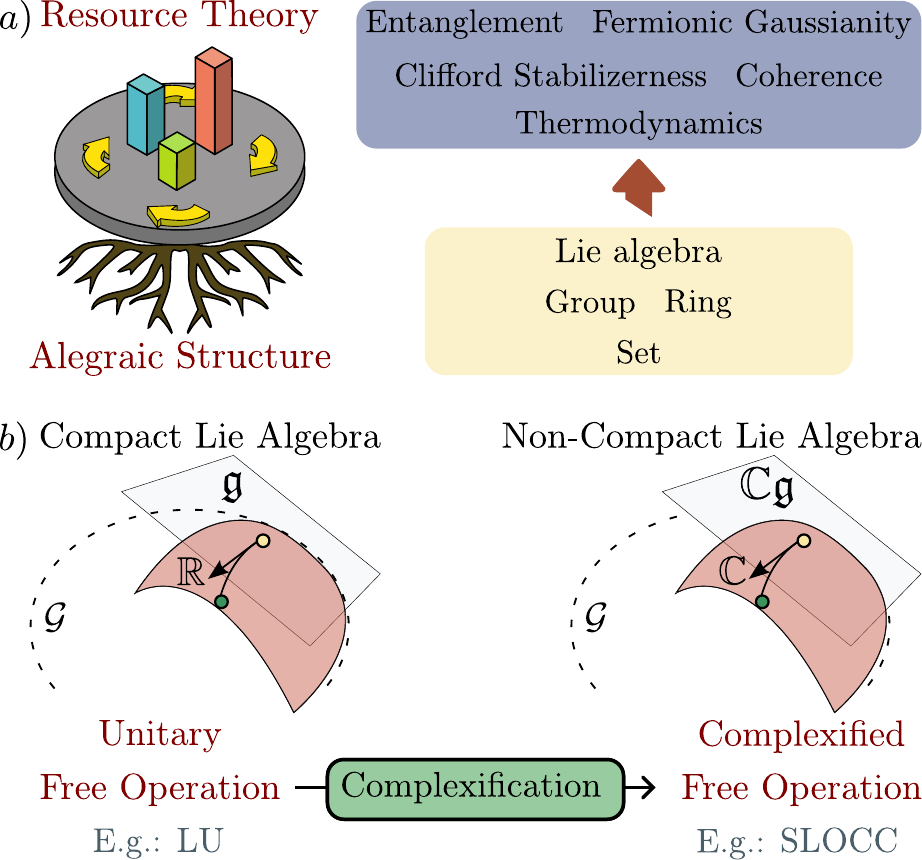}
    \caption{\textbf{Schematic representation of our results.} a) We show that preserved algebraic structures underlie QRTs. We identify such structure for some of the most studied QRTs, finding that it can be a vector space (Lie algebra), group, ring or even just a set. (b) By identifying the complexification as the mechanism which promotes local unitaries (LU) to stochastic local operations and classical communication (SLOCC) in the QRT of entanglement, we introduce a new of set of free operations for Lie group-based QRTs. }
    \label{fig:1}
\end{figure}

In this work, we continue the quest to unify QRTs. In particular, we show that QRTs can be defined in terms of some preferred algebraic structure $\EC$ that must be preserved (see Fig.~\ref{fig:1}). As such, starting from $\EC$, the free operations follow as its automorphisms, which in turn leads to natural notions of free states. Our framework encompasses the results in~\cite{barnum2003generalizations,barnum2004subsystem,viola2007generalized,weinstein2006generalized} as special cases, but also the resource theories in Refs.~\cite{ horodecki2009quantum,plenio2005introduction,beckey2021computable,contreras2019resource,kaur2021resource, garcia2023resource,weedbrook2012gaussian,hebenstreit2019all,diaz2023showcasing,jozsa2008matchgates,mele2024efficient,dias2023classical,goh2023lie,brod2011extending,veitch2014resource,howard2017application,leone2024stabilizer,gour2015resource,goold2016role,lostaglio2019introductory,ng2019resource,hickey2018quantifying,wu2021resource,wu2021operational,horodecki2003reversible,streltsov2018maximal,gour2015resource,luo2019robustness,streltsov2017colloquium,saxena2020dynamical,wu2020quantum,winter2016operational,smith2017quantifying,vaccaro2008tradeoff,gour2008resource,marvian2013theory,marvian2016quantum,martinelli2019quantifying}, for which we explicitly derive $\EC$. Then, we show that our unifying perspective allows us to borrow inspiration from the standard theory of entanglement, and derive a set of resource non-increasing operations for general   Lie-algebraic QRTs, thus answering some of the open problems raised in~\cite{barnum2003generalizations}. \\

\textbf{Unifying framework}.
Let $\HC\cong\mbb{C}^d$ denote some finite dimensional Hilbert space associated to a quantum system of interest, and $\LC(\HC)$ the space of linear operators on $\HC$. Our first result, which leads to a unified approach to QRTs, is the observation that QRTs are characterized by underlying algebraic structures. We define an  
algebraic structure $\EC$ as 
$
\mathcal{E}=\{\mathcal{S},\mathcal{A}\}$
where 
$\mathcal{S}=\{H_\mu\}_{\mu=1}^{\dim(S)}\subseteq\LC(\HC)$ 
is a set of privileged operators and 
$\mathcal{A}$ is a set of algebraic properties. Then we can make the following statement.\vspace{0.2cm}

\noindent\textbf{Claim I}.
\emph{The free operations $\GC$ of a given QRT are those that preserve the algebraic structure, i.e., the automorphisms of $\EC$ through the adjoint action $${\rm Adj}:\GC\subset \GL(\HC)\rightarrow {\rm Aut}(\mathcal{E}) \subset \GL(\LC(\HC))\,.$$
}

From Claim I one can propose different definitions of free states. A  rule of thumb satisfied  throughout (most of) the QRTs studied in this manuscript is: The free states $\FC\subseteq\LC(\HC)$ are those whose orbit is minimal under the free operations~\cite{chitambar2019quantum}. A concrete example corresponds to generalized coherent states, or highest weight states, in the setting of a Lie algebra~\cite{barnum2003generalizations,barnum2004subsystem,perelomov1977generalized,gilmore1974properties,zhang1990coherent} (notably these states minimize generalized uncertainty relations~\cite{delbourgo1977maximum,barnum2004subsystem} and their orbits posses underlying K\"ahler structures~\cite{kostant1982symplectic}, making them ``the most classical''), and convex combinations thereof.  
This also allows one to reinterpret the free states as those that are ``the closest'' to $\EC$. In what follows we will consider several QRTs, and explicitly present the preferred algebraic structure, showcasing how very different QRTs can be treated according to Claim I. \\

\emph{Quantum Thermodynamics.} Let $\HC=\mathbb{C}^d$, and take  
\begin{equation*}
    \SC=\{H\}\,, \quad \AC=\emptyset\,,
\end{equation*}
with $H$ some Hermitian operator in $\LC(\HC)$ which we might identify with a Hamiltonian. Now,   the algebraic structure
$\EC=\{H\}$ 
is simply a \textit{set} and its automorphisms are 
$\GC=\{M\in\mathbb{GL}(\HC)\,|\, [M,H]=0\}\,.$
For instance, if the Hilbert space is composite $\HC=\HC_A\otimes\HC_B$ and $H=H_A\otimes I_B+I_A\otimes H_B$, we recover precisely the setting of the QRT of  quantum thermodynamics \cite{gour2015resource,goold2016role,lostaglio2019introductory,ng2019resource}. \\

\emph{Clifford stabilizerness.} Let $\HC=(\mathbb{C}^2)^n$ be an $n$-qubit Hilbert space, and take 
\begin{equation}
    \SC=\{iI_d\}\cup\{P\}_{P\in\{I_2,X,Y,Z\}^{\otimes n}}\,,\quad \AC=\{\cdot*\cdot \}\,,\nonumber
\end{equation}
so that $\AC$ only contains the matrix multiplication operation. 
The ensuing algebraic structure is that of a \textit{group}, and more specifically, the $n$-qubit Pauli group $P_n$, i.e., $\EC=P_n$.
The automorphisms of the Pauli group can be related to reflection groups (see e.g., ~\cite{planat2010unitary,clark2007generalised}), and only when we restrict to those that fix the center one recovers the Clifford group ${\rm Cl}_n\subset \GC$~\cite{gottesman1998theory,mastel2023clifford,tolar2018clifford,selinger2015generators}\footnote{This is an example where operational and more axiomatic definitions of free operations do not match~\cite{heimendahl2022axiomatic}.}. \\

\emph{Coherence.}  Take $\HC=\mathbb{C}^d$, let
\begin{equation*}
    \SC=\{{\rm diag}(a_1,\ldots a_d)\}_{a_i\in \mathbb{R}},\quad \AC=\{ \cdot+\cdot, \cdot*\cdot \}\,.
\end{equation*}
The ensuing algebraic structure is the \textit{ring} of $d\times d$ diagonal matrices 
$
    \EC=\mathrm{D}_d(\mathbb{R})\cong \mbb{R}^d
$
and the automorphisms $\GC$  are given by the adjoint action of the \textit{generalized permutation matrices} $\mathrm{P}_d(\mbb{R})$. 
Note that $\mathrm{P}_d(\mbb{R})$ is isomorphic to the wreath product group $\mbb{R}\wr S_d \cong \mbb{R}^{ d} \rtimes S_d$, the group with set $\mbb{R}^{ d} \times S_d$ and where $S_d$ acts on $\mbb{R}^{d}$ by permuting the copies of $\mbb{R}$. \\

\emph{Entanglement (LU).} Let us discuss this case in some detail, focusing in the bipartite case. We consider  a Hilbert space $\HC=\HC_A \otimes \HC_B \cong \mathbb{C}^{d_A}\otimes \mathbb{C}^{d_B}$, and take the set of local operators and algebraic properties
\begin{equation}\label{eq:sa-LU}
    \SC=\{iP_A\otimes I_B,I_A\otimes i P_B\}_{P_{A(B)}\in \BC_{A(B)}},\quad     \AC=\{\cdot+\cdot,*_{\mathbb{R}},[\cdot,\cdot]\}\,.\nonumber
\end{equation}
    Above, $I_{A(B)}$ is the identity operator over $\HC_{A(B)}$ and $\BC_{A(B)}$ is a Hermitian set of operators that form a basis of $\LC(\HC_{A(B)})$.
Specifically, in this case $\EC$ is the local unitary Lie algebra 
\begin{equation}
\EC=\mathfrak{u}(\HC_A)\oplus \mathfrak{u}(\HC_B)\,.    
\end{equation}
The inner automorphisms of this \textit{Lie algebra} are given by the adjoint action of the Lie group $\GC=\mathbb{U}(\HC_A)\otimes \mathbb{U}(\HC_B)\subseteq\mathbb{U}(d)$. That is, the free operations correspond to local unitaries (LU), in agreement with the theory of entanglement.  When  $d_A=d_B=\sqrt{d}$ we also have outer automorphism provided by the SWAP operator, so that $\GC=(\mathbb{U}(\sqrt{d})\times \mathbb{U}(\sqrt{d}))\times S_2$. These correspond to the free operations for the  scrambling QRT~\cite{garcia2023resource}, showing that LU entanglement and scrambling collapse under our framework. \\

\emph{Entanglement (SLOCC).} 
Next, we study the case of entanglement via SLOCC~\cite{bennett2000exact,chitambar2014everything,dur2000three,miyake2004multipartite}. We again take $\HC=\HC_A \otimes \HC_B \cong \mathbb{C}^{d_A}\otimes \mathbb{C}^{d_B}$ and consider
\begin{equation}\label{eq:sa-SLOCC} 
    \SC=\{iP_A\otimes I_B,iI_A\otimes P_B\}_{P_{A(B)}\in \BC_{A(B)}},\quad     \AC=\{\cdot+\cdot,*_{\mathbb{C}},[\cdot,\cdot]\}\,.\nonumber
\end{equation}
We remark that the only difference with the previous case is the extension of the field from $\mathbb{R}\to \mathbb{C}$ (see Fig.~\ref{fig:1}).
 It follows that the algebraic structure is the local algebra
\begin{equation}
\EC=    \mathfrak{gl}(\HC_A)\oplus \mathfrak{gl}(\HC_B)=\mathbb{C}\mathfrak{u}(\HC_A)\oplus \mathbb{C}\mathfrak{u}(\HC_B)\,.
\end{equation}
The associated automorphisms are given by $
    \GC=\mathbb{GL}(\HC_A)\times \mathbb{GL}(\HC_B)\supsetneq\mathbb{U}(\HC)$, meaning that the free operations are precisely those in SLOCC~\cite{miyake2004multipartite}\footnote{Again, the gap between LOCC and SLOCC is another example of how an axiomatic approach to free operations differs from an operation one~\cite{koashi2008quantum,duan2009distinguishability}.}. 
    
    Interestingly, we see that while the sets of preferred operators $\SC$ in Eqs.~\eqref{eq:sa-LU} and~\eqref{eq:sa-SLOCC} are the same, a small change in the algebraic operations lead to significantly different free operations (from a compact group of unitaries to resource non-increasing channels of a non-compact group). As we will see below, this simple, yet extremely important observation can serve as inspiration to determine novel resource non-increasing operations in other QRTs.   \\

\emph{Lie algebra.} We now consider the general case when  \begin{equation}\label{eq:sa-La}
    \SC=\{i H_\mu\}_{\mu=1}^{|S|},\quad     \AC=\{[\cdot,\cdot],\cdot+\cdot,*_{\mathbb{R}}\}\,,\nonumber
\end{equation}
where now $H_\mu$ are Hermitian operators in $\LC(\HC)$. Then, the algebraic structure $\EC=\mathfrak{g}$ will form a (representation of a) Lie algebra $\mathfrak{g}$. 
Assuming for simplicity that $\mathfrak{g}$ can be expressed, at most, as a summation of an abelian and a non-abelian ideal, we have that $\GC=e^{\mathfrak{g}}$. 
This setting  captures the QRTs of standard entanglement above, purity~\cite{horodecki2003reversible,streltsov2018maximal,gour2015resource,luo2019robustness}, imaginarity~\cite{hickey2018quantifying,wu2021resource,wu2021operational}\footnote{Note that here free states are not generalized coherent states~\cite{hickey2018quantifying,wu2021resource,wu2021operational,deneris2025analyzing}.}, fermionic Gaussianity~\cite{gigena2015entanglement,weedbrook2012gaussian,hebenstreit2019all,diaz2023showcasing,jozsa2008matchgates,mele2024efficient,dias2023classical,goh2023lie,brod2011extending}, spin coherence~\cite{robert2021coherent,perelomov1977generalized,zhang1990coherent}, and reference frames~\cite{vaccaro2008tradeoff,gour2008resource,marvian2013theory,marvian2016quantum,martinelli2019quantifying}. \\

\textbf{A new class of free operations}.   
An important consequence of our framework, is that we can now import tools from one QRT onto another which, in particular, allows one to gain general insights on what makes a task free. In particular, we now consider 
 a new set of operations inspired by the well-known theory of standard entanglement and the transition from LU to SLOCC. In general, an analogous operational generalization (e.g., add to LU the possibility of performing local measurements, classical communication, etc.) is not feasible as the standard physical intuition  is not available. However, as we showed previously, \textit{our framework identifies the complexification as the fundamental mechanism which generalizes free operations from LU to SLOCC.} Notably, this  insight can be readily exported to any other Lie algebraic structure leading us to introduce the following concept. \vspace{0.2cm}

\noindent\textbf{Definition I (UFOs and CFOs)}.
\emph{Consider a QRT defined by a Lie algebra $\mathfrak{g}$ with unitary free operations (UFO) $\GC_{\rm UFO}=e^{\mathfrak{g}}$. We define the complexified free operations (CFO) as those from $\GC_{\rm CFO}=e^{\mathbb{C}\mathfrak{g}}$, which leads to quantum channels $\Phi(\cdot)=\sum_k M_k \,(\cdot)\, M_k^\dag$ with
$M_k\in e^{\mathbb{C}\mathfrak{g}}$ and $\sum_k M_k^\dag M_k=\id$.}\\
 
As the name suggests, CFOs are genuine free operations. That is, by identifying $\calF$ with the orbit of highest weight states (under UFOs), and statistical mixtures thereof, they map $\calF$ to $\calF$ and hence satisfy the ``free operation postulate'' (see ~\cite{brandao2015reversible}). More formally, we can prove the following result for semi-simple compact Lie algebras (see the Supplemental Material (SM) for a proof of all of our theorems).

\begin{theorem}\label{theo:free-to-free}
   CFOs map free states to free states, i.e., ${\rm Adj}_{\GC_{\rm CFO}}:\FC\rightarrow \FC$ (up to normalization). 
\end{theorem}

Notably, Theorem~\ref{theo:free-to-free} shows how our framework allowed us to obtain genuine free operations from pure mathematical insights. 
More importantly, we remark that CFOs are experimentally accessible, as they can be  implemented by sequential and continuous weak measurements~\cite{jackson2023simultaneous,jackson2023perform}.\\

\textbf{Resource quantifiers and CFOs.} While Theorem~\ref{theo:free-to-free} guarantees the sanity of CFOs as free operations, it does not tell us what happens when a resourceful state is acted upon by such a transformation. Indeed, one can wonder whether CFOs lead to monotonicity properties of UFO resource quantifier\footnote{This is equivalent to asking, e.g., how the purity of a reduced density matrix--an LU invariant--changes under SLOCC.}.  Answering this question will require QRT and witness specific analysis, but we here provide strong evidence suggesting a positive answer to this question. Specifically,  in Lie-algebra-based QRTs, one can use the so-called $\mathfrak{g}$-purity~\cite{barnum2003generalizations,ragone2023unified,bermejo2025characterizing} as a resource quantifier: $$\mathcal{P}(\rho):=\frac{1}{\mathcal{N}_\mathfrak{g}}\sum_{i=1}^{\dim(\mathfrak{g})} \Tr[\rho g_i]^2\,,$$ where $\{ig_i\}_i$ forms an orthogonal basis of $\mathfrak{g}$, and where $\mathcal{N}_\mathfrak{g}$ is the dimension of the Cartan subalgebra of $\mathfrak{g}$, so that  $\mathcal{P}(\rho)=1$ iff $\rho=\dya{\psi} \in \FC$. Indeed, the  $\mathfrak{g}$-Purity is, by definition, invariant under UFOs and smaller values of it indicate more resource. For instance, in the QRT of LU entanglement, the associated $\mathfrak{g}$-Purity is proportional to the average standard purity of all marginal states. The latter is clearly invariant under LU but only non-increasing on average under SLOCC. As such, we now seek to study how the $\mathfrak{g}$-Purity changes under CFOs. For simplicity, we will focus on two very different and important QRTs: spin coherence, and fermionic Gaussianity.

For convenience, we recall that in the QRT of spin coherence we  take $\HC=\mathbb{C}^d$, and $\EC$ the spin $s=(d-1)/2$ irreducible representation of $\mathfrak{su}(2)={\rm span}_{\mathbb{R}}\{iJ_x,iJ_y,iJ_z\}$. Then, in the QRT of $n$ spinless fermions we set $\HC=2^n$, and $\EC$ is the spinor representation of $\mathfrak{so}(2n)$ obtained from the product of two Majorana operators (i.e., $\mathfrak{so}(2n)={\rm span}_{\mathbb{R}}\{c_\mu c_\nu\}$ with $1\leq \mu<\nu\leq 2n$)~\cite{diaz2023showcasing}. We begin with the following theorem for the QRT of spin coherence.

\begin{figure}[t!]
    \centering
    \includegraphics[width=.8\linewidth]{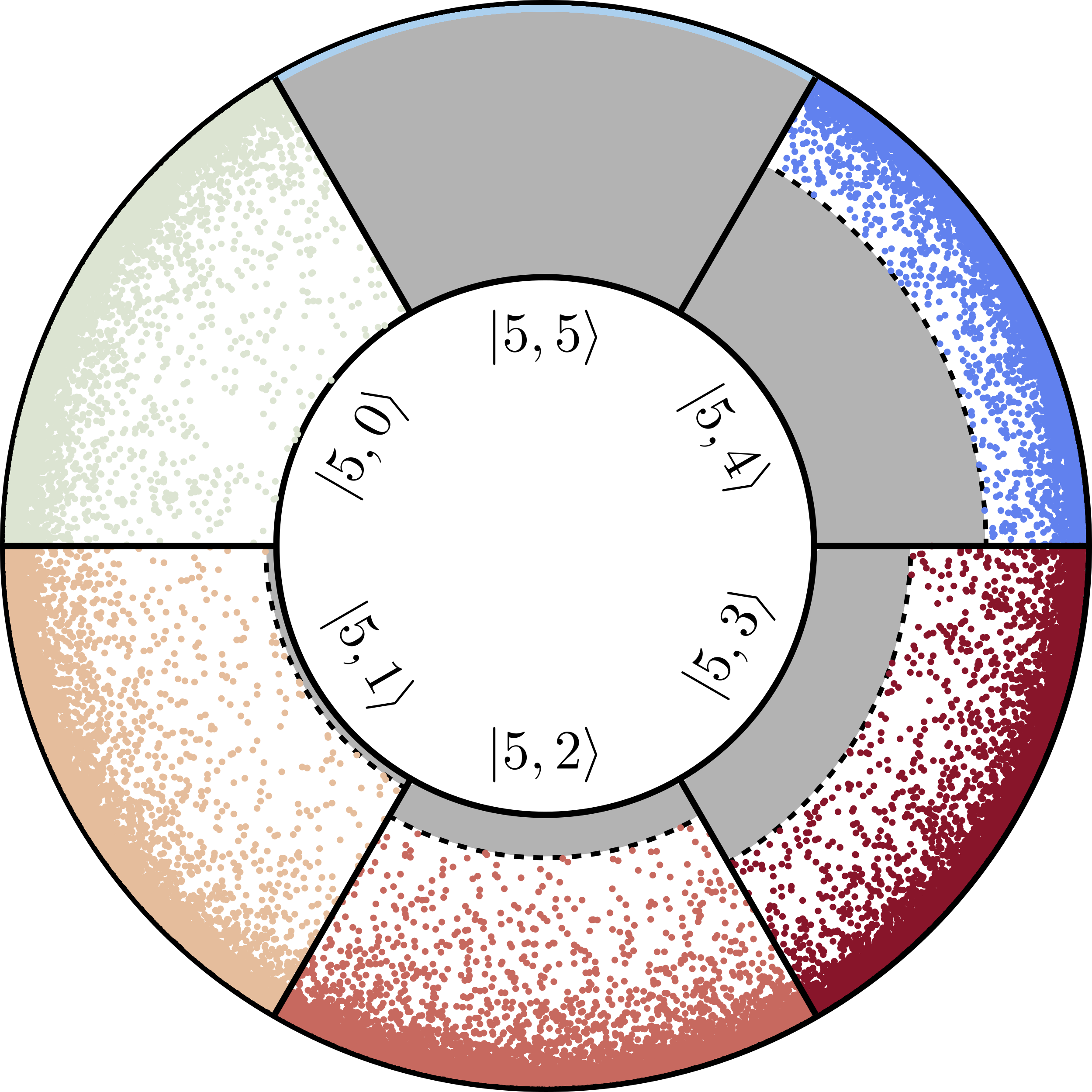}
    \caption{\textbf{Change in $\mathfrak{g}$-Purity under CFOs for the QRT of spin coherence.} Here we show the algebraic purity for $10^4$ (normalized) states $M|s,m\rangle$ with $M\in\GC_{\rm CFO}=\mathbb{GL}(2)$ for $s=5$ and for the indicated  values of $m$. The inner (outer) circle corresponds to $\mathfrak{g}$-Purity equal to zero (one). As per Theorem~\ref{th:sm}, the action of $M$ can only strictly decrease the resourcefulness--increase the $\mathfrak{g}$-Purity~\cite{bermejo2025characterizing}-- of the state with respect to the  starting value of $m^2/s^2$ (dashed lines in each quadrant). Moreover, $\ket{s,s}$ is mapped to other free states, further showcasing  Theorem~\ref{theo:free-to-free}.}
    \label{fig:weightstates}
\end{figure}

\begin{figure}[t!]
    \centering
    \includegraphics[width=1\linewidth]{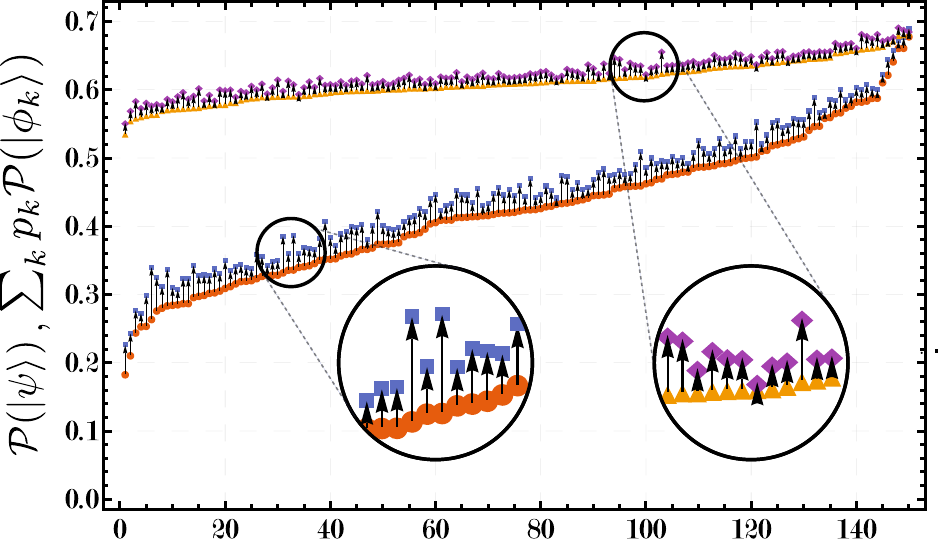}
    \caption{\textbf{Numerical evidence of conjecture 1.} The red circles (orange triangles) represent the $\mathfrak{su}(2)$-purity ($\mathfrak{so}(2n)$-purity) of 150 initial random states--ordered by their purity--in QRT of spin coherence for $s=5$ (fermionic Gaussianity for $n=8$ qubits). The blue squares (purple diamonds) are the average purity  after acting with a CFO channel on these states. Arrows show the direction of change in the purity (i.e., CFOs only decrease on average the resourcefulness). The circles show zoomed-in sections of the plot as indicated.  }
    \label{fig:average}
\end{figure}

\begin{theorem}\label{th:sm}
   Let  $|s,m\rangle$ denote the weight states of $\mathfrak{su}(2)$, such that $\PC(\ket{s,m})=m^2/s^2$ (i.e., all weight states except for $m=\pm s$ are resourceful). Then, the transformations in  $\GC_{\rm CFO}=\mathbb{GL}(2)=e^{\mathbb{C}\mathfrak{su}(2)}$ are resource non-increasing as 
   \begin{equation}
       \PC\left(\frac{M\ket{s,m}}{\sqrt{\bra{s,m}M\ad M\ket{s,m}}} \right)\geq \PC(\ket{s,m})\,, \quad \forall M\in \mathbb{GL}(2)\nonumber\,,
   \end{equation}
 with equality holding if $m=s$ as per Theorem~\ref{theo:free-to-free}.  
\end{theorem}

\noindent In Fig.~\ref{fig:weightstates} we show an application of the previous result. 

Importantly, the strict inequality of Theorem~\ref{th:sm} can be used to draw a parallel with the QRT of bipartite entanglement where the weight states are separable states with Schmidt's rank of one, a quantity that cannot increase, not even probabilistically, under standard SLOCC \cite{dur2000three}. Extending even further the analogy with SLOCC, where for arbitrary states entanglement can increase with some probability but not on average, we introduce the following conjecture. \vspace{0.2cm}

\noindent\textbf{Conjecture 1}.
\emph{Given a Lie algebra-based QRT, the $\mathfrak{g}$-purity does not decrease on average under a CFO channel of the form $\Phi\,(|\psi\rangle)=\sum_k p_k |\phi_k\rangle\langle \phi_k|$, where  $M_k\in e^{\mathbb{C}\mathfrak{g}}$,   $p_k=\langle \psi|M_k^\dag M_k|\psi\rangle$, and $|\phi_k\rangle=\frac{M_k |\psi\rangle}{\sqrt{\langle \psi|M_k^\dag M_k|\psi\rangle}}$. That is, } 
\begin{equation*}
 \sum_k p_k\mathcal{P}(\ket{\phi_k})\geq \PC(\ket{\psi})\,.
\end{equation*}

Clearly, Theorem \ref{th:sm} implies the validity of the conjecture for weight states of $\mathfrak{su}(2)$.
In Fig.~\ref{fig:average} we provide numerical evidence for arbitrary states of both the spin coherence ($\mathfrak{su}(2)$) and fermionic Gaussianity ($\mathfrak{so}(2n)$) QRTs.    We refer the reader to the SM for additional details. \\

\textbf{Discussions.} In this work we present a simple realization: QRTs can be defined from, and relative to, a preferred underlying algebraic structure that must be preserved. More than a mathematical artifact, such result has profound implications as it allows us to take lessons from one QRT and apply them onto another. In particular, we have illustrated this fact by obtaining non-unitary resource non-increasing CFOs operations in Lie group-based QRT, thus recovering the results in~\cite{barnum2003generalizations}. Importantly, the fact that these new free operations can be implemented via sequential, continuous, isotropic weak measurement~\cite{jackson2023simultaneous,jackson2023perform}, indicates a deep connection between free operations and schemes for optimal state tomography~\cite{shojaee2018optimal}.

Moreover, our work raises the importance of further characterizing all the free operations which arise as the automorphisms of standard algebraic structures. For instance, in the QRT of Clifford stabilizerness we find that the Pauli group automorphisms (which contain non-unitary maps)  are much larger than those contained in the Clifford group. Indeed, it would be extremely interesting to see if other notions of free operations in stabilizerness QRT~\cite{veitch2014resource,heimendahl2022axiomatic} match exactly the automorphisms of the Pauli group or if there still exists gaps between those sets. More generally, we also open the question as to whether one can define free operations from more general transformations of the preferred algebraic structure (e.g., endomorphisms). In particular, if one wants to account for transformations that connect different sized-Hilbert spaces, then one likely needs to use morphisms and more abstract categorical transformations~\cite{coecke2015categorical}. 

As such, the burden is now on the community to show that existing measures of resource (which are invariant under unitary transformation), are either strictly non-increasing, or non-increasing in average, under our new set of operations. Further, we also pave the way towards CFO invariant quantities, which could reveal different and non-equivalent forms of resources in general QRTs (just like three qubit states can be entangled in two inequivalent ways~\cite{dur2000three}). 
\vspace{0.12cm}

\textbf{Acknowledgments.\!} We thank David Wierichs, Nathan Killoran and Diego Garc\'ia-Martin for useful discussions. The research presented in this article  was supported by the National Security Education Center Informational Science and Technology Institute (ISTI) using the Laboratory Directed Research and Development (LDRD) program of Los Alamos National Laboratory (LANL) project number 20240479CR-IST. NLD was supported by  the Center for Nonlinear Studies at LANL. AAM was supported by the U.S. Department of Energy (DOE) through a quantum computing program sponsored by the LANL ISTI and by  the German Federal Ministry for Education and Research (BMBF) under the project FermiQP.  P. Braccia, P. Bermejo and AED acknowledge support by the LDRD program of LANL under project numbers 20230049DR and 20230527ECR. P. Bermejo also acknowledges constant support from DIPC. ML and MC were supported by  LANL ASC Beyond Moore’s Law project, and by the Quantum Science Center, a National Quantum Information Science Research Center of the U.S. DOE.

%

\clearpage
\newpage

\onecolumngrid

\section*   {Supplemental Information for ``A unified approach to quantum resource theories and a new class of free operations'' }

\setcounter{theorem}{0}

\section{Proof of Theorems}
\subsection{Theorem 1}
Here we prove Theorem I which states that complexified free operations (CFOs) map free states to free states in Lie algebra-based QRTs. For convenience we recall that given a QRT whose unitary free operations (UFO) are obtained as $\GC=e^{\mathfrak{g}}$, for a semi-simple compact Lie algebra $\mathfrak{g}$, we define the set of free states $\FC$ as the orbit of the highest weight under $\GC$, and convex combinations thereof. Then, defining the CFOs as $\GC_{\rm CFO}=e^{\mathbb{C}\mathfrak{g}}$, we intend to show that
\begin{theorem}\label{theo:free-to-free-si}
   CFOs map free states to free states, i.e., ${\rm Adj}_{\GC_{\rm CFO}}:\FC\rightarrow \FC$ (up to normalization). 
\end{theorem}

Before presenting the main proof let us first introduce a useful Lemma. \\

\noindent\textbf{Supplemental Lemma 1}. 
\emph{
For a semi-simple compact Lie algebra we can write 
       any $M\in e^{\mathbb{C}\mathfrak{g}}$ as 
     \begin{equation}\label{eq:supllemmaiwa}
         M=Ue^{\sum_i \alpha_i h_i}e^{\sum_j x_j e^+_j}\,,
     \end{equation}
     where $U\in e^{\mathfrak{g}}$ is a unitary in the set of CFOs, $\mathfrak{g}=\text{span}\{h_i,e^+_i,e^-_i\}$ the Cartan-Weyl decomposition for $\mathfrak{g}$, and with $\alpha\in \mathbb{R}$, $x_j\in \mathbb{C}$.
}

  \begin{proof}
   The Lemma relies on obtaining a convenient Iwasawa decomposition of the complexified algebra. 
    The first step is to notice that $\mathbb{C}\mathfrak{g}=\mathfrak{g}\oplus i\mathfrak{g}$ is a Cartan decomposition of the complexified algebra. As a matter of fact, $[\mathfrak{g},\mathfrak{g}]\subset \mathfrak{g}$, $[i\mathfrak{g},i\mathfrak{g}]\subset \mathfrak{g}$, $[i\mathfrak{g},\mathfrak{g}]\subset i\mathfrak{g}$. Notice that we can associate this decomposition to the involution $\theta(X)=-X^\dag$.

    Then, according to the Iwasawa decomposition we can write $\mathbb{C}\mathfrak{g}=\mathfrak{g}\oplus i\mathfrak{h}\oplus \mathfrak{n}$ where we are using that $i\mathfrak{h}$ is the maximal abelian subalgebra of $i\mathfrak{g}$ and with $\mathfrak{n}=\text{span}\{e^+_i\}$ for $\{h_i,e^+_i,e^-_i\}$ a Cartan-Weyl basis of $\mathfrak{g}$. Notice that $\mathfrak{n}$ is defined over the field of complex numbers as we are considering the complexified algebra whose Cartan-Weyl basis is simply given by the same set $\{h_i,e^+_i,e^-_i\}$ but over $\mathbb{C}$.  The Iwasawa decomposition of the algebra induces the Iwasawa decomposition of the group thus leading to the statement of the Lemma.  
  \end{proof}

We are now in a position to prove {Theorem~\ref{theo:free-to-free-si}}. \begin{proof}
   Consider a pure free state associated with a given algebra $\mathfrak{g}$. We can always write it as a free operation on the highest weight state $|HW\rangle$. Then, without loss of generality we can think of any pure free state as $|HW\rangle$ with corresponding Cartan decomposition with basis $\{h_i,e^+_i,e^-_i\}$. By means of the Supplemental Lemma 1 we can write any $M\in e^{\mathbb{C}\mathfrak{g}}$ as $M=Ue^{\sum_i \alpha_i h_i}e^{\sum_j x_j e^+_j}$. This leads to 
\begin{equation}
    M|HW\rangle=Ue^{\sum_i \alpha_i h_i}|HW\rangle=\lambda U|HW\rangle
\end{equation}
with $\lambda$ a complex constant.
Then under adjoint action we have
\begin{equation}
    M|HW\rangle\langle HW|M^\dag=|\lambda|^2 U|HW\rangle\langle HW|U^\dag 
\end{equation}
which is a generalized coherent state up to normalization $|\lambda|^2$. By linearity, the same holds for any $\rho\in \FC$. 
\end{proof}
${}$\\

Next, let us briefly analyze quantum channels where all the Kraus operators are CFOs. Here, we can show that  
$\Phi_{\text{CFO}}\,(\rho)=\sigma$ with $\rho,\sigma \in \calF$ by using the ideas in the proof of Theorem 1.
As a matter of fact, when $M_k\in e^{\mathbb{C}\mathfrak{g}}$ we have
\begin{equation}
    \Phi\,(|HW\rangle)=\sum_k M_k |HW\rangle\langle HW| M_k^\dag=\sum_k |\lambda_k| ^2 U_k|HW\rangle\langle HW|U_k^\dag\,, 
\end{equation}
with $\sum_k |\lambda_k| ^2=1$
for trace preserving quantum operations where $\sum_k M_k^\dag M_k=1$. 
It is then clear that $\Phi\,(|HW\rangle)\in \calF$. By linearity, the same holds if we apply the channel to any $\rho\in \calF$. In summary, $\Phi(\rho)\in \FC$ for $\rho \in \FC$.

Notice also that for a pure state and $U_k=U$ for all $k$ (in the basis where the pure state is the highest weight state), the channel maps free pure states to free pure states. In particular, for $U=\id$, namely, all the channels are generated by $i\mathfrak{g}$,  the pure free states are a fixed point of the CFO channel.

\subsection{Theorem 2}
Here we provide a proof for  Theorem 2, which we restate for convenience. 

\begin{theorem}\label{th:sm-si}
   Let  $|s,m\rangle$ denote the weight states of $\mathfrak{su}(2)$, such that $\PC(\ket{s,m})=m^2/s^2$ (i.e., all weight states except for $m=\pm s$ are resourceful). Then, the transformations in  $\GC_{\rm CFO}=\mathbb{GL}(2)=e^{\mathbb{C}\mathfrak{su}(2)}$ are resource non-increasing as 
   \begin{equation}
       \PC\left(\frac{M\ket{s,m}}{\sqrt{\bra{s,m}M\ad M\ket{s,m}}} \right)\geq \PC(\ket{s,m})\,, \quad \forall M\in \mathbb{GL}(2)\nonumber\,,
   \end{equation}
 with equality holding if $m=s$ as per Theorem~\ref{theo:free-to-free-si}.  
\end{theorem}

Our strategy is to first provide a closed-form expression for the purity and then bound it.

Let us first recall a few useful facts. 
As stated in the main text, the following quantity measures the resourcefulness of a state with respect to the QRT for $\mathfrak{g}=\mathfrak{su}(2)$~\cite{barnum2003generalizations,ragone2023unified,bermejo2025characterizing}
\begin{equation}
    \PC(\ket{\psi})=\frac{1}{s^2}\sum_{\mu=x,y,z}\bra{\psi}J_\mu\ket{\psi}^2\,.
\end{equation} 
For instance, one can show that $\PC(\ket{s,m})=m^2/s^2$ which is clearly maximized and equal to one for the highest and lowest weight states, or spin coherent states, $m=\pm s$.
In what follows, we are interested in taking an element $M\in\mathbb{SL}(2)=e^{\mathbb{C}\mathfrak{g}}$ for $\mathfrak{g}=\mathfrak{su}(2)$. In particular, we can use Eq.\ \eqref{eq:supllemmaiwa} from the Supplemental Lemma 1 (Iwasawa decomposition) to express this operator as
\begin{equation}
    M=e^{i\theta J_y}e^{\alpha J_z}e^{\eta J_+}\,,
\end{equation}
with $\alpha,\theta\in \mathbb{R}$, and $\eta \in \mathbb{C}$. 
Given that $\PC(U\ket{s,m})=\PC(\ket{s,m})$ for any $U\in\mathbb{SU}(2)\subseteq\mathbb{SL}(2)$, then we can take, without loss of generality $M=e^{\alpha J_z}e^{\eta J_+}$ and omit the first rotation.  After an application of the transformation $|\psi\rangle\to |\psi'\rangle= \frac{M|\psi\rangle}{\sqrt{\langle \psi| M^\dag M |\psi\rangle}}$ the purity is modified as follows: 
\begin{equation}
\begin{split}
     \PC(\ket{\psi'})&=\frac{1}{s^2 \langle \psi| M^\dag M |\psi\rangle^2} \sum_{\mu=x,y,z}\bra{\psi}e^{\eta^\ast J_-}e^{\alpha J_z}J_\mu e^{\alpha J_z}e^{\eta J_+}\ket{\psi}^2\,,
\end{split}
\end{equation} 
with $\langle \psi| M^\dag M |\psi\rangle=\bra{\psi}e^{\eta^\ast J_-}e^{\alpha J_z} e^{\alpha J_z}e^{\eta J_+}\ket{\psi}$. \\

\emph{Closed form of the purity of weight states under CFOs}.
Let us first assume that the state is the highest weight state $|\psi\rangle=|s,s\rangle$. In this case $M|s,s\rangle=e^{\alpha s}|s,s\rangle$ so that $|\psi'\rangle=|\psi\rangle$, i.e., the state is unchanged. Clearly this implies that the purity is preserved as well in agreement with Theorem~\ref{theo:free-to-free-si}. Consider now other weight states $|\psi\rangle=|s,m\rangle$ with $0\leq m < s$ (the case of negative $m$ follows by symmetry).

 Let us now recall that 
\begin{align}\label{eq:J+-on-sm}
    J_\pm\ket{s,m}=c(s,m)\ket{s,m\pm1}\,,
\end{align}
with $c(s,m)=\sqrt{s(s+1)-m(m+1)}$. 
Then, if we want to compute the action
\begin{align}
    e^{\eta J_+}\ket{s,m}&=\sum_{k=0}^\infty \frac{\eta^k}{k!}J_+^k \ket{s,m}=\sum_{k=0}^{s-m} \frac{\eta^k}{k!}J_+^k \ket{s,m}\,,
\end{align}
we can use Eq.~\eqref{eq:J+-on-sm} to find
\begin{equation}
  J_+^k \ket{s,m}=\zeta(s,m,k)  \ket{s,m+k}\,.
\end{equation}
Here,  we defined the coefficients $\zeta(s,m,k)\in\mathbb{R}$ as
\begin{align}
\begin{split}
\zeta(s,m,k)\!=\!\sqrt{(s-m)(s+m+1) (s-m-1)_{k-1}(2+m+s)_{k-1}}\,,\nonumber  
\end{split}
\end{align}
and we  assumed that $k\leq s$. Above $(a)_k$ denotes the Pochhammer symbol given by 
\begin{equation}
    (a)_k=a(a+1)\cdots (a+k-1)=\frac{\Gamma(a+k)}{\Gamma(a)}\,.
\end{equation}
Combining the previous results we find that $\ket{\phi}=e^{\alpha J_z}e^{\eta J_+}\ket{s,m}$ can be expanded as 
\begin{align}
    \ket{\phi}=e^{\alpha m}\ket{s,m}+\sum_{k=1}^{s-m} \frac{\eta^k}{k!}e^{\alpha (m+k)}\zeta(s,m,k) \ket{s,m+k}\nonumber\,.
\end{align}

Then,  the normalization coefficient is
\begin{align}\label{supeq:norm}
\bra{\phi}\phi\rangle&=e^{2\alpha m}\left(1+\sum_{k=1}^{s-m} \frac{|\eta|^{k}}{(k!)^2}e^{2\alpha k}\zeta(s,m,k)^2\right)=e^{2m\alpha}{}_2F_1[m-s,s+m+1;1;-|\eta|^2e^{-2\alpha}]\,,
\end{align}
with $_2F_1[a,b;c;d]$ the hypergeometric function.
Next, we want to evaluate the expectation values $\bra{\phi}J_\mu|\phi\rangle$ for $\mu\in\{x,y,z\}$. We find 
\begin{align}\label{supeq:Z}
\bra{\phi}J_z|\phi\rangle&=e^{2\alpha m}\left(m+\sum_{k=1}^{s-m} \frac{|\eta|^{k}}{(k!)^2}e^{2\alpha k}\zeta(s,m,k)^2(m+k)\right)=\frac{1}{2}\frac{\partial}{\partial \alpha}\bra{\phi}\phi\rangle\,.
\end{align}
Then, we need to evaluate
\begin{align}\label{supeq:XY}
\bra{\phi}J_x|\phi\rangle=\frac{1}{2}\bra{\phi}(J_++J_-)|\phi\rangle\,,\quad \bra{\phi}J_y|\phi\rangle=\frac{1}{2i}\bra{\phi}(J_+-J_-)|\phi\rangle\,.
\end{align}
In order to do so, we use the fact that 
\begin{align}
    J_+\ket{\phi}=&e^{\alpha m}c(s,m)\ket{s,m+1}+\sum_{k=1}^{s-m-1} c(s,m+k) \frac{\eta^k}{k!}e^{\alpha (m+k)}\zeta(s,m,k) \ket{s,m+k+1}\,,\nonumber
\end{align}
while
$
    \bra{\phi}J_-\ket{\phi}=(\bra{\phi}J_+\ket{\phi})^*\,.
$
After some simple algebra, this leads to the closed-form expression
\small
\begin{align}\label{supeq:Jplus}
    \bra{\phi}J_+\ket{\phi}=&e^{\alpha (2m+1)}c(s,m)\eta^\ast \zeta(s,m,1)+\sum_{k=1}^{s-m-1} c(s,m+k) \frac{\eta^\ast|\eta|^{2k}}{k!(k+1)!}e^{\alpha (2m+2k+1)}\zeta(s,m,k)\zeta(s,m,k+1)\nonumber\\
   =& \frac{\eta^\ast  e^{\alpha +2 \alpha  m} (m-s) (m+s+1) \left(e^{2 \alpha } |\eta| ^2 m \, _2F_1\left(m-s+1,m+s+2;3;-e^{2 \alpha } |\eta| ^2\right)-\, _2F_1\left(m-s,m+s+1;2;-e^{2 \alpha } |\eta| ^2\right)\right)}{e^{2 \alpha } |\eta|^2+1}\,.
\end{align}\\
\normalsize

Combining Eqs.~\eqref{supeq:norm}--\eqref{supeq:Jplus}, and setting $\ket{\psi'}=\frac{M\ket{s,m}}{\sqrt{\bra{s,m}M\ad M\ket{s,m}}}$  we finally obtain
\small
\begin{equation}
\begin{split}
     \!\!\PC(\ket{\psi'})&=\frac{\left(m \, _2F_1\left(m-s,m+s+1;1;-e^{2 \alpha } |\eta| ^2\right)-e^{2 \alpha } |\eta| ^2 (m-s) (m+s+1) \, _2F_1\left(m-s+1,m+s+2;2;-e^{2 \alpha } |\eta| ^2\right)\right){}^2}{s^2 \, _2F_1\left(m-s,m+s+1;1;-e^{2 \alpha } |\eta| ^2\right){}^2}\\&+\frac{e^{2 \alpha } |\eta| ^2 (m-s)^2 (m+s+1)^2 \left(_2F_1\left(m-s,m+s+1;2;-e^{2 \alpha } |\eta| ^2\right)-e^{2 \alpha } |\eta|^2 m \, _2F_1\left(m-s+1,m+s+2;3;-e^{2 \alpha } |\eta| ^2\right)\right){}^2}{s^2 \, _2F_1\left(m-s,m+s+1;1;-e^{2 \alpha } |\eta| ^2\right){}^2\left(e^{2 \alpha } |\eta| ^2+1\right)^2}\,.
\end{split}
\end{equation}
\normalsize

\begin{figure}[t!]
    \centering
    \includegraphics[width=0.55\linewidth]{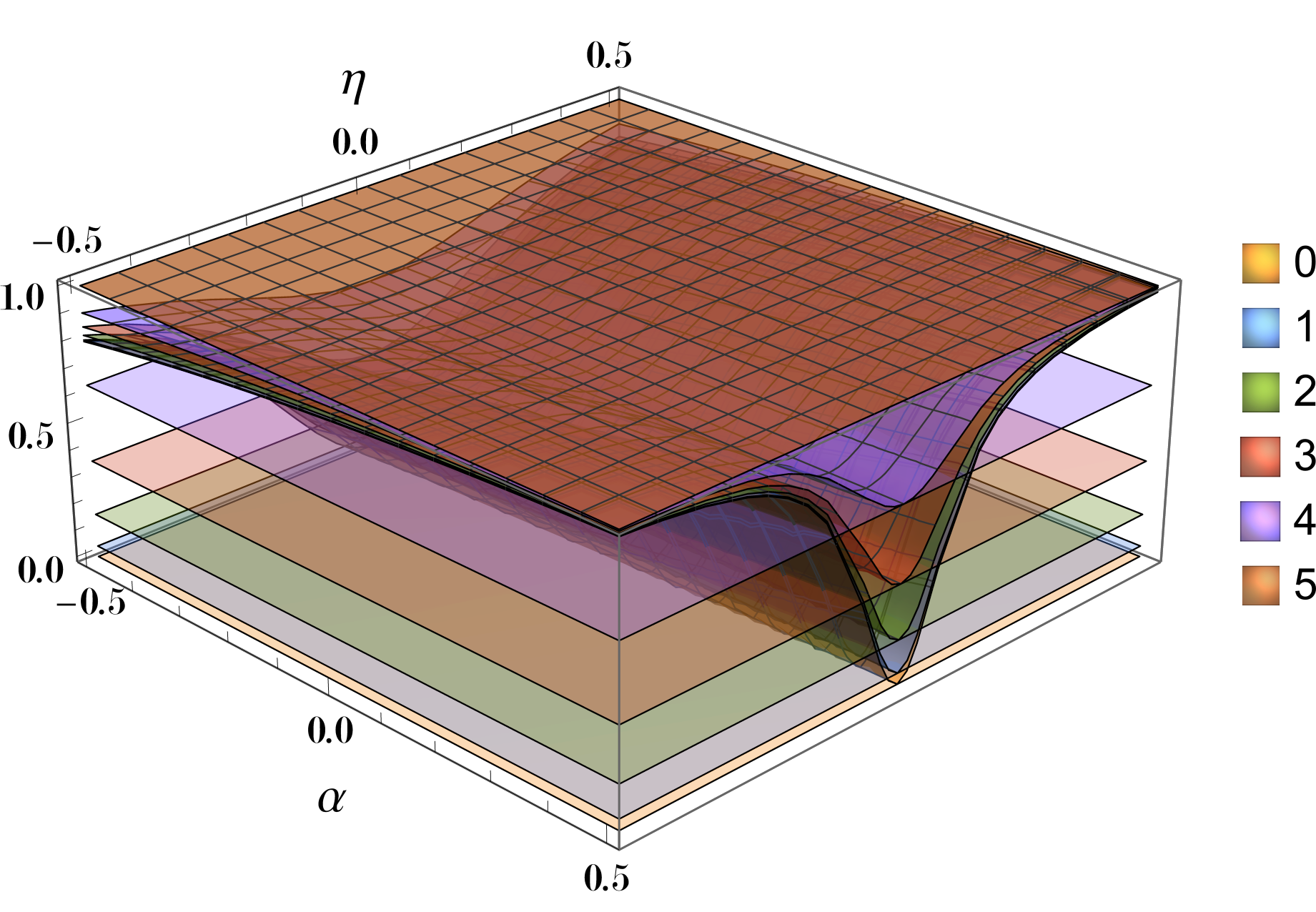}
    \caption{\textbf{Purity of different weight states after acting with a CFO parameterized by $\alpha,\eta$.} Here $s=5$ and $\eta$ is taken as real (equivalent to considering the absolute value, see Eq.\ \eqref{eq:purityclosed}). The curves correspond to $\PC(|\psi'\rangle)$ while the planes correspond to $\PC(|s,m\rangle)$ with the two matching for $\eta=0$. The different colors indicate the different values of $m$, with $m=s$ leading to purity one, unaffected by CFOs, and $m=0$ the most resourceful state. }
    \label{fig:purity}
\end{figure}

We can slightly simplify this expression   by defining $z=-e^{2\alpha}|\eta|^2$ and using basic properties of the hypergeometric functions, so that 
\begin{equation}\label{eq:purityclosed}
\begin{split}
     \PC(\ket{\psi'})=\left(\frac{m}{s}+\frac{1}{s}z\frac{d}{dz} \log[_2F_1\left(m-s,m+s+1;1;z\right)]\right)^2
     -z \,[G(m,s,z)]^2\,,
\end{split}
\end{equation}
with $G(m,s,z):=\frac{(m-s) (m+s+1) \left(\, _2F_1\left(m-s,m+s+1;2;z\right)+z m \, _2F_1\left(m-s+1,m+s+2;3;z\right)\right)}{s \, _2F_1\left(m-s,m+s+1;1;z\right)\left(1-z\right)}$ a real function.\\

In Figure \ref{fig:purity} we show $\PC(|\psi'\rangle)$ for different CFOs and compare it with $\PC(|s,m\rangle)$. The behavior agrees with Theorem~\ref{th:sm-si}.
We can now use the closed form of Eq.\ \eqref{eq:purityclosed} to prove {Theorem~\ref{th:sm-si}}.

\begin{proof}

Consider Eq.\ \eqref{eq:purityclosed}. Notice that $z\leq 0$ meaning that the second term is always positive (or null) since  $G(m,s,z)\in \mathbb{R}$.

On the other hand, we can easily show that $$z\frac{d}{dz} \log[_2F_1\left(m-s,m+s+1;1;z\right)]\geq 0\,.$$
As a matter of fact, $$z\frac{d}{dz} \log[_2F_1\left(m-s,m+s+1;1;z\right)]=z(m-s)(m+s+1)\frac{_2F_1\left(m-s+1,m+s+2;2;z\right)}{_2F_1\left(m-s,m+s+1;1;z\right)}\,,$$
with $z(m-s)(m+s+1)\geq 0$ so we only need to show that $\frac{_2F_1\left(m-s+1,m+s+2;2;z\right)}{_2F_1\left(m-s,m+s+1;1;z\right)}\geq 0$ for $z\leq 0$. This follows by noting that both numerator and denominator are positive. Let us consider the denominator first. By definition we can write  \begin{align*}
    _2F_1\left(m-s,m+s+1;1;z\right)&=\sum_k (m-s)_k(m+s+1)_k\frac{z^k}{k!^2}=\sum_k (-1)^k \binom{s-m}{k}(m+s+1)_k \frac{z^k}{k!}\\&=\sum_k  \binom{s-m}{k}(m+s+1)_k \frac{|z^k|}{k!}\geq 0
\end{align*} with $(-1)^kz^k=|z^k|$ as $z^k$ is negative for $k$ odd and where we used that $m-s$ is a negative integer. An analogous expansion holds for the numerator leading to $_2F_1\left(m-s+1,m+s+2;2;z\right)=\sum_k  \binom{s-m-1}{k}(m+s+2)_k \frac{|z^k|}{(k+1)!}\geq 0$.

With this information we can simply write 
\begin{equation*}
    \begin{split}
         \PC(\ket{\psi'})&=\left[\frac{m}{s}+\frac{1}{s}z\frac{d}{dz} \log[_2F_1\left(m-s,m+s+1;1;z\right)]\right]^2
     -z [G(m,s,z)]^2\geq \left[\frac{m}{s}+\frac{1}{s}z\frac{d}{dz} \log[_2F_1\left(m-s,m+s+1;1;z\right)]\right]^2\\&\geq \frac{m^2}{s^2}=\PC(\ket{s,m})\,.
    \end{split}
\end{equation*}
In summary, we have proven that $\PC(\ket{\psi'})\geq \PC(\ket{s,m})$ for any $z$ and then for any CFO.

 \end{proof}

It is interesting to notice that the statement of  Theorem~\ref{th:sm-si} holds for the following intuitive reason (revealed by the calculations and the proof): The purity of weight states comes from the Cartan operator $J_z$ alone as the other components vanish. Thus under CFOs the components $J_x$, $J_y$ can only increase the purity (the term $-zG(s,m,z)^2$). At the same time,
when acting with a CFO upon a weight state, the purity of $J_z$ also increases.

\section{Numerical implementation of CFOs and details on the conjecture}

Here we discuss the details of our numerical evidence on the main text conjecture. The latter states that the resources, measured by the $\mathfrak{g}$-purity, are non-increasing on average. Mathematically, this means that given Kraus operators  $M_k\in e^{\mathbb{C}\mathfrak{g}}$, and a pure state $\ket{\psi}$ we find that 
\begin{equation}
  \sum_k p_k\calP(\ket{\phi_k}) 
    \geq \calP(\ket{\psi})\,,
\end{equation}
for $p_k=\langle \psi|M_k^\dag M_k|\psi\rangle$, and $|\phi_k\rangle=\frac{M_k |\psi\rangle}{\sqrt{\langle \psi|M_k^\dag M_k|\psi\rangle}}$.
Let us also recall that more purity indicates less resources.

In our numerical test we considered Kraus operators of the form
\begin{equation}
    M_{k_1k_2\dots k_N}:=2^{-N/2}\Big[\cos\Big(\epsilon \sum_i h_i g_i\Big)-(-1)^{k_N}\sin\Big(\epsilon \sum_i h_i g_i\Big)\Big]\dots\Big[\cos\Big(\epsilon \sum_i h_i g_i\Big)-(-1)^{k_1}\sin\Big(\epsilon \sum_i h_i g_i\Big)\Big]\,,
\end{equation}
where $k_t=0,1$. Here $ig_i\in \mathfrak{g}$ with $g_i$ hermitian and $h_i$ real coefficients taken as $h_i\in (-1,1)$. One can easily verify the completeness relation $\sum_{k_1,k_2,\dots , k_N} M^\dag_{k_1k_2\dots k_N} M_{k_1k_2\dots k_N}=\id$. The previous equation implies that one can obtain the associated quantum process by means of ancillas and unitary evolution \cite{nielsen2000quantum}. For instance, one might use a single ancilla of $N$ qubits once or sequentially couple the system to a qubit $N$ times.
For $\epsilon$ small we can think of the latter procedure as sequential weak measurements. In this regime we can write 
$M_{k_1k_2\dots k_N}=E_{k_1k_2\dots k_N}+\mathcal{O}(T \epsilon)$, for  $T=\epsilon N$ and with 
\begin{equation}
    E_{k_1k_2\dots k_N}:=2^{-N/2}\exp[\epsilon \sum_{t=1}^N (-1)^{k_t}\sum_i h_i g_i]\in e^{\mathbb{C}\mathfrak{g}}\,.
\end{equation}
In particular $E_{\textbf{0}}=2^{-N/2}\exp[T \sum_i h_i g_i]$, showing that we can reach arbitrary regions of $\mathbb{C}G$ with arbitrary precision by increasing $N$ at fixed $T$. 
One can easily add unitary parts to the Kraus operators, but since we know the purity is invariant under such transformations we simply focus in the non-unitary sector.

In the Figure 3 of the main text, we considered Kraus operators $M_{k_1k_2k_3k_4k_5}$ corresponding to $\epsilon=0.02$ and $N=5$, namely $2^5$ Kraus operators for both algebras, with corresponding $g_i$ each. For each point in the plot, the initial state is chosen at random and the coefficients $h_i$ as well. In the case of $SU(2)$ we considered $s=5$ as in Figure 2. For $SO(2n)$ we considered $n=8$ fermionic modes.

\end{document}